\documentclass[aps,prd,twocolumn,10pt,nofootinbib,longbibliography]{revtex4-1}
\usepackage{amssymb}
\usepackage{graphicx}
\usepackage{amsmath}
\usepackage{hyperref}
\usepackage{subfigure}
\usepackage{multirow}
\usepackage{setspace}
\usepackage{verbatim}
\usepackage{float}
\usepackage{color}
\usepackage{ulem}
\usepackage{bm}
\usepackage[utf8]{inputenc}

\begin{document}

\title{Searching for double-peak and doubly broken gravitational-wave spectra from Advanced LIGO-Virgo's first three observing runs}

\author{Wang-Wei Yu$^{1,2}$}
\email{yuwangwei@mail.itp.ac.cn}
\author{Shao-Jiang Wang$^{1}$}
\email[Corresponding author:~]{schwang@itp.ac.cn}

\affiliation{$^1$CAS Key Laboratory of Theoretical Physics, Institute of Theoretical Physics, Chinese Academy of Sciences, Beijing 100190, China}
\affiliation{$^2$School of Physical Sciences, University of Chinese Academy of Sciences (UCAS), Beijing 100049, China}

\begin{abstract}
The current LIGO-Virgo observing run has been pushing the sensitivity limit to touch the stochastic gravitational-wave backgrounds (SGWBs). However, no significant detection has been reported to date for any single dominated source of SGWBs with a single broken-power-law (BPL) spectrum. Nevertheless, it could equally well escape from existing Bayesian searches from, for example, two comparable dominated sources with two separate BPL spectra (double-peak case) or a single source with its power-law behavior in the spectrum broken twice (doubly broken case). In this paper, we put constraints on these two cases but specifically for the model with cosmological first-order phase transitions from Advanced LIGO-Virgo's first three observing runs. We found strong negative evidence for the double-peak case and hence place 95\% C.L. upper limits $\Omega_\mathrm{BPL,1}<5.8\times10^{-8}$ and $\Omega_\mathrm{BPL,2}<4.4\times10^{-8}$ on the two BPL spectra amplitudes with respect to the unresolved compact binary coalescence (CBC) amplitude $\Omega_\mathrm{CBC}<5.6\times10^{-9}$. We further found weak negative evidence for the doubly broken case and hence place 95\% C.L. upper limit $\Omega_\mathrm{DB}<1.2\times10^{-7}$ on the overall amplitude of the doubly broken spectrum with respect to  $\Omega_\mathrm{CBC}<6.0\times10^{-9}$. In particular, the results from the double-peak case have marginally ruled out the strong super-cooling first-order phase transitions at LIGO-Virgo band.

\end{abstract}

\maketitle

\section{Introduction}

The sensitivity limits have been persistently pushed forward during the first three observing runs (O1~\cite{LIGOScientific:2018mvr}, O2~\cite{LIGOScientific:2020ibl}, and O3~\cite{LIGOScientific:2021djp}) of the Advanced LIGO~\cite{LIGOScientific:2014pky} and Advanced Virgo~\cite{VIRGO:2014yos} gravitational wave (GW) detectors, which might uncover the stochastic GW backgrounds (SGWBs)~\cite{Christensen:2018iqi,Renzini:2022alw,vanRemortel:2022fkb} from unresolved sources of both astrophysical and cosmological origins. The unresolved sources of astrophysical origins mainly consist of the compact binary coalescences (CBCs) from unresolved individual sources such as binary black hole and neutron star mergers~\cite{Rosado:2011kv,Zhu:2011bd,Marassi:2011si,Wu:2011ac,Zhu:2012xw} as well as other more exotic sources that are also more difficult to be observed, including the core-collapse supernovae~\cite{Buonanno:2004tp,Sandick:2006sm,Marassi:2009ib,Zhu:2010af}, rotating neutron stars \cite{Ferrari:1998jf,Howell:2010hm,Zhu:2011pt,Marassi:2010wj,Rosado:2012bk,Wu:2013xfa,Lasky:2013jfa}, stellar-core collapses~\cite{Crocker:2015taa,Crocker:2017agi,Finkel:2021zgf}, and boson clouds around black holes~\cite{Brito:2017wnc,Brito:2017zvb,Fan:2017cfw,Tsukada:2018mbp,Palomba:2019vxe,Sun:2019mqb,KAGRA:2021tse}, to name just a few.

The SGWBs of cosmological origins~\cite{Caprini:2018mtu} can embrace much more rich physics~\cite{Cai:2017cbj,Bian:2021ini}. Primordial GWs produced from an inflationary era~\cite{Turner:1996ck} uniquely mark the energy scale of cosmic inflation, in particular, the scalar-induced secondary GWs~\cite{Nakamura:2004rm,Ananda:2006af,Osano:2006ew,Baumann:2007zm,Cai:2018dig} during inflation depict the curvature perturbations at small scales. The SGWBs from cosmological first-order phase transitions (FOPTs)~\cite{Caprini:2015zlo,Mazumdar:2018dfl,Caprini:2019egz,Hindmarsh:2020hop,Caldwell:2022qsj} and cosmic strings~\cite{Kibble:1976sj,Sarangi:2002yt,Damour:2004kw,Siemens:2006yp} necessarily encode the new physics beyond the standard model of particle physics, while the SGWBs from primordial black hole (PBH) mergers~\cite{Wang:2016ana} can constrain the PBH abundance in the dark matter. However, multiple sources of these SGWBs of cosmological origins can be equally well present simultaneously in the GW data.

The detectability for SGWBs below the confusion limit is by no doubt difficult compared to the individually resolvable GW events that make up a tiny fraction of all GW signals present in the detector time stream. The most recent isotropic search~\cite{KAGRA:2021kbb} from O3 combined with previous ones from O1~\cite{LIGOScientific:2016jlg} and O2~\cite{LIGOScientific:2019vic} is consistent with uncorrelated noise and hence places upper limits on the normalized GW energy density for power-law spectral-index values of 0 (flat), 2/3 (CBCs), and 3 (causality). Other efforts of searches with LIGO-Virgo Collaboration for the SGWBs from cosmic strings~\cite{LIGOScientific:2017ikf,LIGOScientific:2021nrg}, FOPTs~\cite{Romero:2021kby,Huang:2021rrk,Jiang:2022mzt,Badger:2022nwo}, and induced GWs~\cite{Romero-Rodriguez:2021aws,Mu:2022dku} all return a null result.

However, a simple Bayesian search with a single broken power-law (BPL) spectrum for any single but dominated source of SGWBs might just miss possible detections on, for example, two comparable dominated sources of SGWBs with two separate single-BPL spectra (double-peak case) or a single source of SGWBs with its power-law behavior in the spectrum broken twice (double-broken case). The double-peak (DP) case also includes a single source of SGWBs but already with two peaks by nature, for example, two-step FOPT~\cite{Bigazzi:2020avc,Zhao:2022cnn}, one-step FOPT but with comparable GWs from both wall collisions and sound waves when bubbles collide during the transition to a near constant terminal wall velocity~\cite{Cai:2020djd}, induced GWs with two peaks for some particular configuration on curvature perturbations~\cite{Cai:2019amo}, oscillons with cuspy potentials~\cite{Liu:2017hua} during preheating era. The doubly broken (DB) case can be found in an analytic evaluation on the GWs from wall collisions beyond the envelope approximation~\cite{Jinno:2017fby} and a hybrid simulation for the sound waves~\cite{Jinno:2020eqg} (see also~\cite{Jinno:2022mie}).

In this paper, we search for the SGWB signals in the cases with DP and DB spectra from Advanced LIGO-Virgo's first three observing runs but with a special focus on SGWBs from FOPTs. The models are introduced in Sec.~\ref{sec:Models} and constrained in Sec.~\ref{sec:DataAnalysis}, and the results are summarized in Sec.~\ref{sec:Results}. 

\section{Models}\label{sec:Models}

\begin{figure}
\centering
\includegraphics[width=0.48\textwidth]{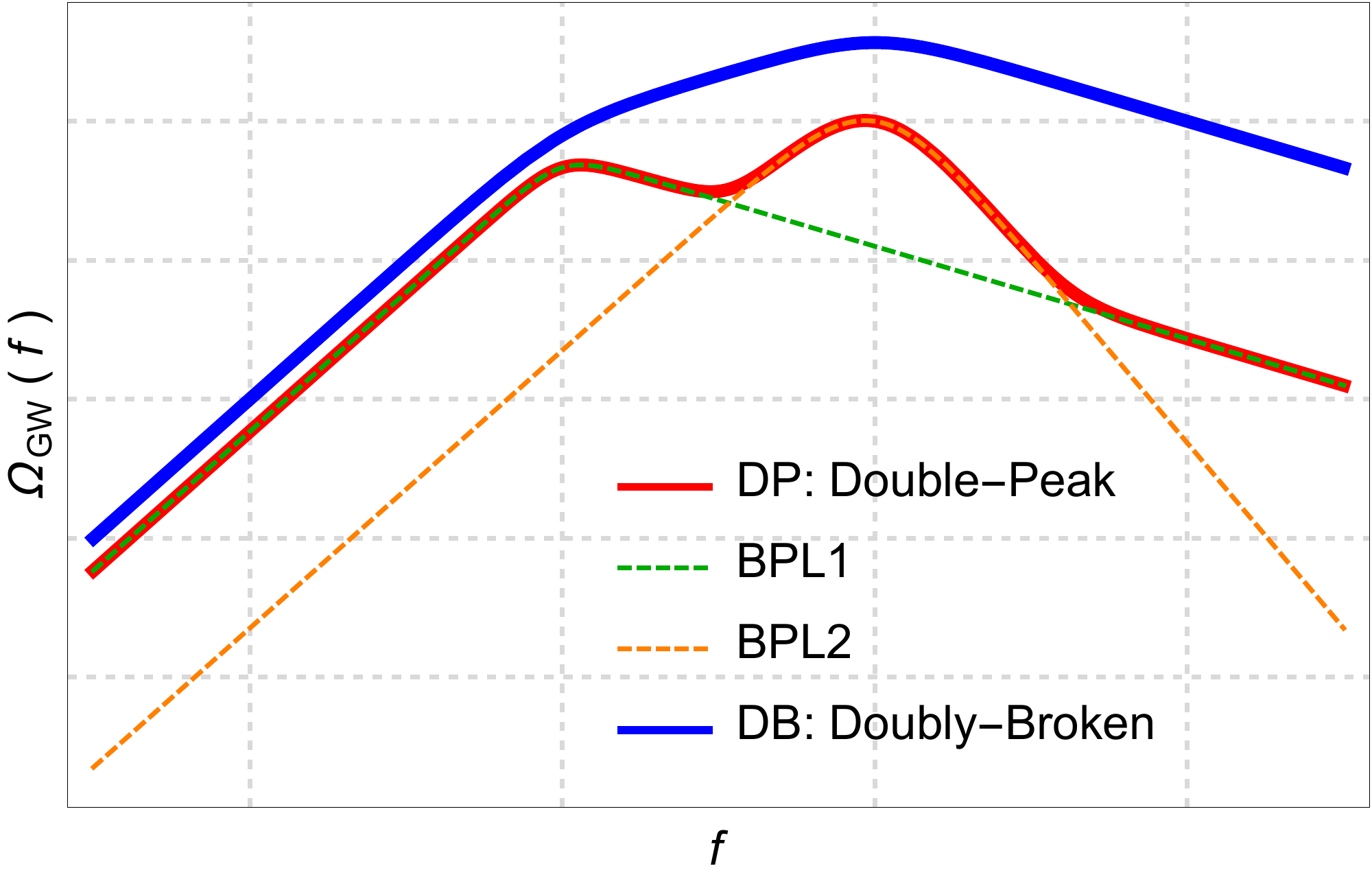}\\
\caption{The schematic comparison between DP (red solid) and DB (blue solid) spectra for the SGWB. The DP spectrum consists of two separate BPL spectra (green and orange dashed) while the DB spectrum admits a three-section power-law scaling.}\label{fig:schematic}
\end{figure}

The BPL spectrum of SGWBs can be modeled as
\begin{equation}\label{eq:BPL}
    \Omega_\mathrm{BPL}(f; \bm{\theta}) = \Omega_* \left( \frac{f}{f_*} \right)^{n_1} \left[\frac{1+(f/f_*)^\Delta}{2}\right]^{(n_2 - n_1) / \Delta}
\end{equation}
with $\bm{\theta}\equiv(\Omega_*, f_*, n_1, n_2, \Delta)$, where $\Omega_*$ is the peak amplitude at the peak frequency $f_*$, $n_1$ ($=3$ for causality) and $n_2$ are the asymptotic slopes on the far left and far right ends of the peak frequency, respectively, and $1/\Delta$ is the peak transition width. We define the DP spectrum as simply the sum of two separate BPL spectra,
\begin{align}\label{eq:TBPL}
\Omega_\mathrm{DP}(f;\{\bm{\theta}_1,\bm{\theta}_2\})=\Omega_\mathrm{BPL}(f;\bm{\theta}_1)+\Omega_\mathrm{BPL}(f;\bm{\theta}_2)
\end{align}
with $\bm{\theta}_i\equiv(\Omega_{*,i}, f_{*,i}, n_{1,i}, n_{2,i}, \Delta_i)$ for the peaks $i=1,2$,
while the DB spectrum is modeled as a three-section power-law scaling by
\begin{align}\label{eq:DBPL}
    \Omega_\mathrm{DB}&(f;\bm{\theta}) = \frac{\Omega_*}{\left(\frac{f}{f_l}\right)^{-n_l} + \left(\frac{f}{f_l}\right)^{-n_m} + \left(\frac{f_h}{f_l}\right)^{-n_m}\left(\frac{f}{f_h}\right)^{-n_h}} \nonumber\\
    & \propto \left\{ \begin{aligned}
        &(f/f_l)^{n_l}, & f &\ll f_l, \\
        &(f/f_l)^{n_m}, & f_l \ll & f \ll f_h, \\
        (f_h/ & f_l)^{n_m}(f/f_h)^{n_h}, & f_h & \ll f,
    \end{aligned} \right.
\end{align}
with $\bm{\theta}\equiv(\Omega_*,f_l,f_h,n_l,n_m,n_h)$, where we will assume $n_l > n_m > n_h$ and $f_l < f_h$ with a specific example~\cite{Jinno:2020eqg} in mind. An explicit comparison between DP and DB spectra is shown in Fig.~\ref{fig:schematic}.
The reference SGWBs from unresolved CBCs can be approximated by a $f^{2/3}$ power-law spectrum~\cite{Callister:2016ewt} in the inspiral phase as
\begin{equation}\label{eq:CBC}
    \Omega_\mathrm{CBC} (f;\bm{\theta}) = \Omega_\mathrm{ref}\left( \frac{f}{f_\mathrm{ref}} \right)^{2/3}
\end{equation}
with $\bm{\theta}\equiv(\Omega_\mathrm{ref},f_\mathrm{ref})$,  where $\Omega_\mathrm{ref}$ is the reference amplitude at the reference frequency $f_\mathrm{ref}$ that will be fixed at $25$ Hz around the most sensitive frequency band of the LIGO-Virgo network. For all the above models, the uncorrelated Gaussian noise is implicitly included with $\Omega_\mathrm{N}(f)=0$.

The primary motivation to test above models beyond the simple BPL model comes from SGWBs from FOPTs, which usually occur for breaking some continuous symmetry that would form a potential barrier for the false vacuum decaying into the true vacuum. The vacuum decay process proceeds via spontaneous nucleations of true vacuum bubbles in the false vacuum plasma, followed by rapid expansions of bubble walls until violent collisions, along with which the expanding bubble walls also stimulate fluid motions of the thermal plasma. Therefore, both the bubble wall collisions and plasma fluid motions would generate SGWBs. The simple BPL spectrum with $n_1=3$, $n_2=-1$, and $\Delta=4$ can depicts the GW spectrum from bubble-wall collisions under the dubbed envelope approximation~\cite{Jinno:2016vai}, in which case the overlapping parts of thin walls are neglected upon collisions. However, by going beyond the envelope approximation, the analytic modeling~\cite{Jinno:2017fby} of wall collisions reveals a three-section power-law scaling with $n_l=3$, $n_m=1$, and $n_h=-1$ that can be described by a DB spectrum. Furthermore, the GWs from plasma fluid motions especially the dominated contributions from sound waves can also be fitted by the simple BPL spectrum with $n_1=3$, $n_2=-4$, and $\Delta=2$ as suggested by numerical simulations~\cite{Hindmarsh:2013xza,Hindmarsh:2015qta,Hindmarsh:2017gnf}.  However, analytic modelings~\cite{Hindmarsh:2016lnk,Hindmarsh:2019phv,Cai:2023guc} seem to prefer a DB spectrum with $n_l=3$, $n_m=1$, and $n_h=-3$ but still with some uncertainty in determining its high-frequency slope within $-3\leq n_h\leq-1$~\cite{Cai:2023guc}. Nevertheless, we will stick to the fitting spectrum from numerical simulation instead of the analytic estimation for the sound waves. On the other hand, it is probable for some specific particle physics model of FOPT in its particular parameter space that the contributions from the envelope collisions and sound waves are comparable, which can be described by a DP spectrum.

\section{Data Analysis}\label{sec:DataAnalysis}

\begin{table}
\small
    \centering
    \caption{The prior choices for the combined SGWB models BPL+CBC, DP+CBC, and DB+CBC.}
    \begin{tabular}{c|c|c}
    \hline
    \hline
    \multicolumn{3}{c}{\textbf{BPL + CBC}}\\
    \hline
        Parameter  &  \multicolumn{2}{c}{Prior}\\
    \hline
        $\Omega_\mathrm{ref}$ &  \multicolumn{2}{c}{LogUniform $(10^{-10}, 10^{-7})$ }\\
    \hline
        $\Omega_\mathrm{*}$ & \multicolumn{2}{c}{LogUniform $(10^{-9},10^{-4})$ } \\
    \hline
        $f_*$ &  \multicolumn{2}{c}{Uniform $(0,256)$ Hz }  \\
    \hline
        $n_1$ &   \multicolumn{2}{c}{3} \\
    \hline
       & sound waves & envelope-wall collisions  \\
    \hline
        $n_2$ & -4 &  -1  \\
    \hline
        $\Delta$  & 2  & 4 \\
    \hline
    \hline
    \multicolumn{3}{c}{\textbf{DP + CBC}}\\
    \hline
        Parameter &   \multicolumn{2}{c}{Prior} \\
    \hline
        $\Omega_\mathrm{ref}$ &  \multicolumn{2}{c}{LogUniform $(10^{-10}, 10^{-7})$ } \\
    \hline
        $\Omega_{*,i}$ &   \multicolumn{2}{c}{LogUniform $(10^{-9}, 10^{-4})$ } \\
    \hline
        $f_{*,1}$ &  \multicolumn{2}{c}{Uniform $(0,256)$ Hz } \\
    \hline
        $f_{*,2} (>f_{*,1})$ &  \multicolumn{2}{c}{Uniform $(0,256)$ Hz} \\
    \hline
        $n_{1,i}$ &   \multicolumn{2}{c}{3}\\
    \hline
        $n_{2,1}$ & -1  & envelope-wall collisions\\
    \hline
       $n_{2,2}$ &  -4 & sound waves \\
    \hline
        $\Delta_1$ & 4  & envelope-wall collisions\\
    \hline
        $\Delta_2$ & 2  & sound waves\\
    \hline
    \hline
    \multicolumn{3}{c}{\textbf{DB + CBC}}\\
    \hline
        Parameter &   \multicolumn{2}{c}{Prior}\\
    \hline
        $\Omega_\mathrm{ref}$ &   \multicolumn{2}{c}{LogUniform $(10^{-10}, 10^{-7})$ } \\
    \hline
        $\Omega_\mathrm{*}$ &  \multicolumn{2}{c}{LogUniform $(10^{-9}, 10^{-4})$}  \\
    \hline
        $f_l$ &  \multicolumn{2}{c}{Uniform $(0,256)$ Hz}  \\
    \hline
        $f_h (>f_l)$ &   \multicolumn{2}{c}{Uniform $(0,256)$ Hz}  \\
    \hline
        $n_l (>n_m)$ & 3 & \multirow{3}{*}{beyond envelope wall collisions}  \\
     \cline{1-2}
        $n_m (>n_h)$ & 1  & \\
    \cline{1-2}
        $n_h$ &  -1  & \\
    \hline
    \hline
    \end{tabular}
    \label{tab:priors}
\end{table}

We closely follow the method outlined in Refs.~\cite{KAGRA:2021kbb,Romero:2021kby,Romero-Rodriguez:2021aws} to search for SGWBs in the current GW data. The log-likelihood for the model parameter set $\bm{\theta}$ is estimated by~\cite{Mandic:2012pj,Callister:2017ocg,Meyers:2020qrb,Matas:2020roi}
\begin{align}
\log p(\hat{C}_{IJ}|\bm{\theta},\lambda)\propto-\frac12\sum_k\frac{\left[\hat{C}_{IJ}(f_k)-\lambda\Omega_\mathrm{GW}(f_k;\bm{\theta})\right]^2}{\sigma_{IJ}^2(f_k)}
\end{align}
with the sum $k$ running over the frequency bins, where $\hat{C}_{IJ}(f)$ is the cross-correlation statistic for the baseline $IJ$ with $I, J={H,L,V}$ for the LIGO-Hanford, LIGO-Livingston, and Virgo (HLV) detectors, $\sigma^2_{IJ}(f)$ is the variance of $\hat{C}_{IJ}(f)$ in the small signal-to-noise ratio limit~\cite{Allen:1997ad}, and $\lambda$ accounts for the calibration uncertainties of the detectors~\cite{Sun:2020wke} that would be eventually marginalized over~\cite{Whelan:2012ur}. 
Since the current data still favors a pure Gaussian noise model~\cite{KAGRA:2021kbb}, the contribution from Schumann resonances is negligible \cite{Meyers:2020qrb,Thrane:2013npa,Coughlin:2018tjc}.
The final likelihood is obtained by summing over multiple log-likelihoods for different baselines in order to constrain the model parameters. As the SGWB from CBCs is an indispensable part of any SGWBs at the LIGO-Virgo band, we will search for SGWBs specifically from FOPTs for the combined models BPL+CBC, DP+CBC, and DB+CBC with their parameter priors depicted in Table.~\ref{tab:priors}.
For model comparison, we adopt the ratios of evidence $\log \mathcal{B}_\mathrm{Noise}^{i + \mathrm{CBC}}$ and $\log \mathcal{B}_\mathrm{CBC}^{i + \mathrm{CBC}}$ from the Bayes factor to evaluate the preference for a specific SGWB model over a pure Gaussian noise model and a CBC background, respectively.

\begin{figure*}
    \centering
    \includegraphics[width=0.8\textwidth]{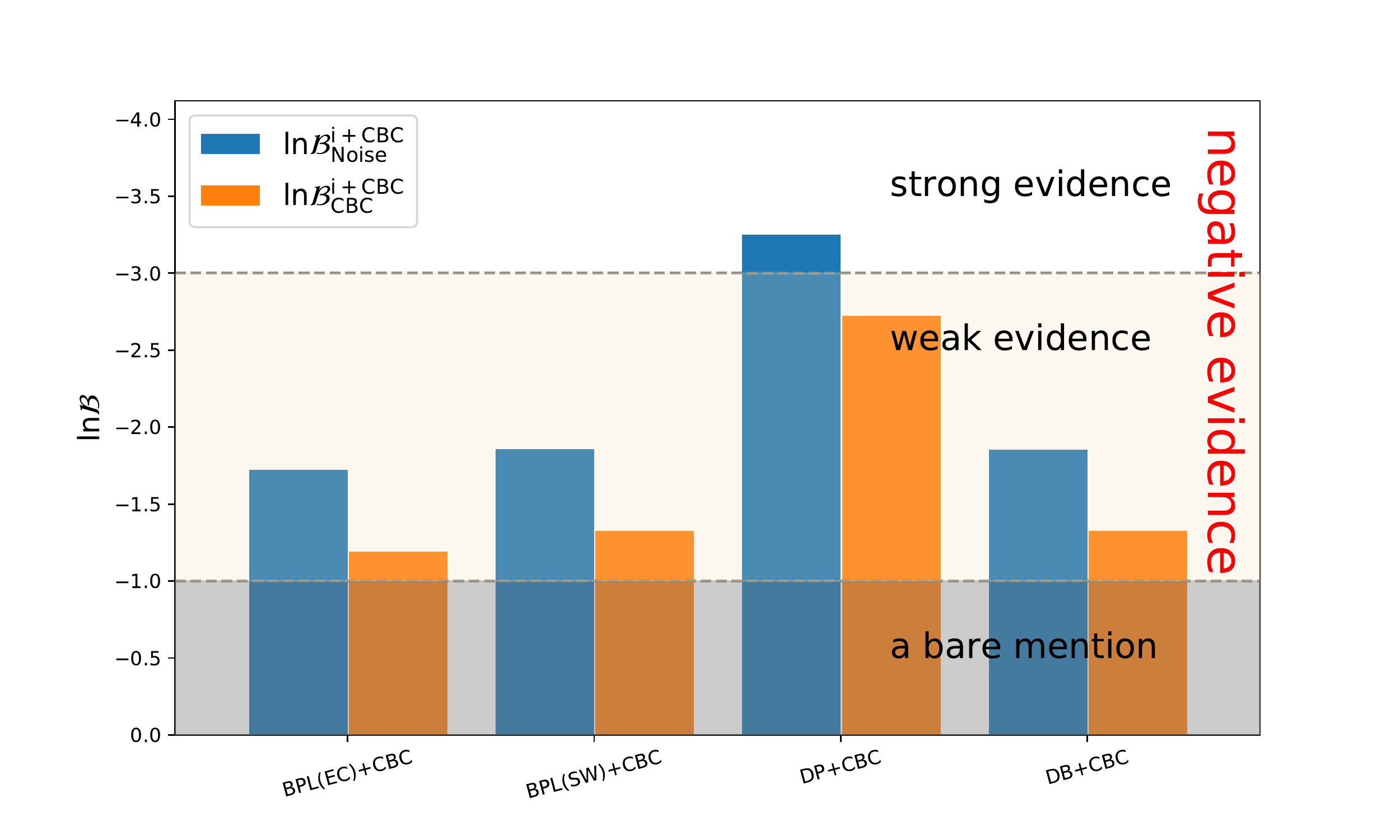}
    \caption{The Bayes ratios for model comparisons among BPL+CBC models with priors fixed by envelope collisions (EC) and sound waves (SW), DP+CBC model (envelope collisions + sound waves), and DB+CBC model (wall collisions beyond envelope approximation).}
    \label{fig:BeyesFactor}
\end{figure*}

\section{Results}\label{sec:Results}

The Bayes-ratio comparison of models constrained by the dynamic nested sampling package \texttt{dynesty} \cite{Speagle:2019ivv} in BILBY~\cite{Ashton:2018jfp} is summarized in Fig.~\ref{fig:BeyesFactor}, which will be described in details shortly below. The general conclusion is that, the SGWB from a  DP spectrum is even more disfavored than the SGWB with a single BPL spectrum or a DB spectrum, but all of which are not detected compared to the backgrounds from either uncorrelated Gaussian noises or CBCs.

\subsection{The BPL+CBC model} 

To compare with previous results in the literature, we repeat the BPL model of Ref.~\cite{Romero:2021kby} but with an extra $2$ factor in \eqref{eq:BPL} so that $\Omega_*$ is exactly the peak amplitude at the peak frequency $f_*$, while in Ref.~\cite{Romero:2021kby} the peak amplitude is actually $\Omega_\mathrm{BPL}(f_*)=2^{(n_2-n_1)/\Delta}\Omega_*$ instead of $\Omega_*$. For $n_2<0<n_1$ and $\Delta>0$, the upper bound $\Omega_*=5.6\times10^{-7}$ obtained in Ref.~\cite{Romero:2021kby} with fixed $n_1=3$ and $\Delta=2$ actually overestimates the true peak amplitude with a larger overestimation for steeper slopes (larger $|n_1|$ and/or $|n_2|$) or a wider transition width (a smaller $\Delta$) around the peak position. 

\begin{figure*}
    \centering
    \includegraphics[width=0.49\textwidth]{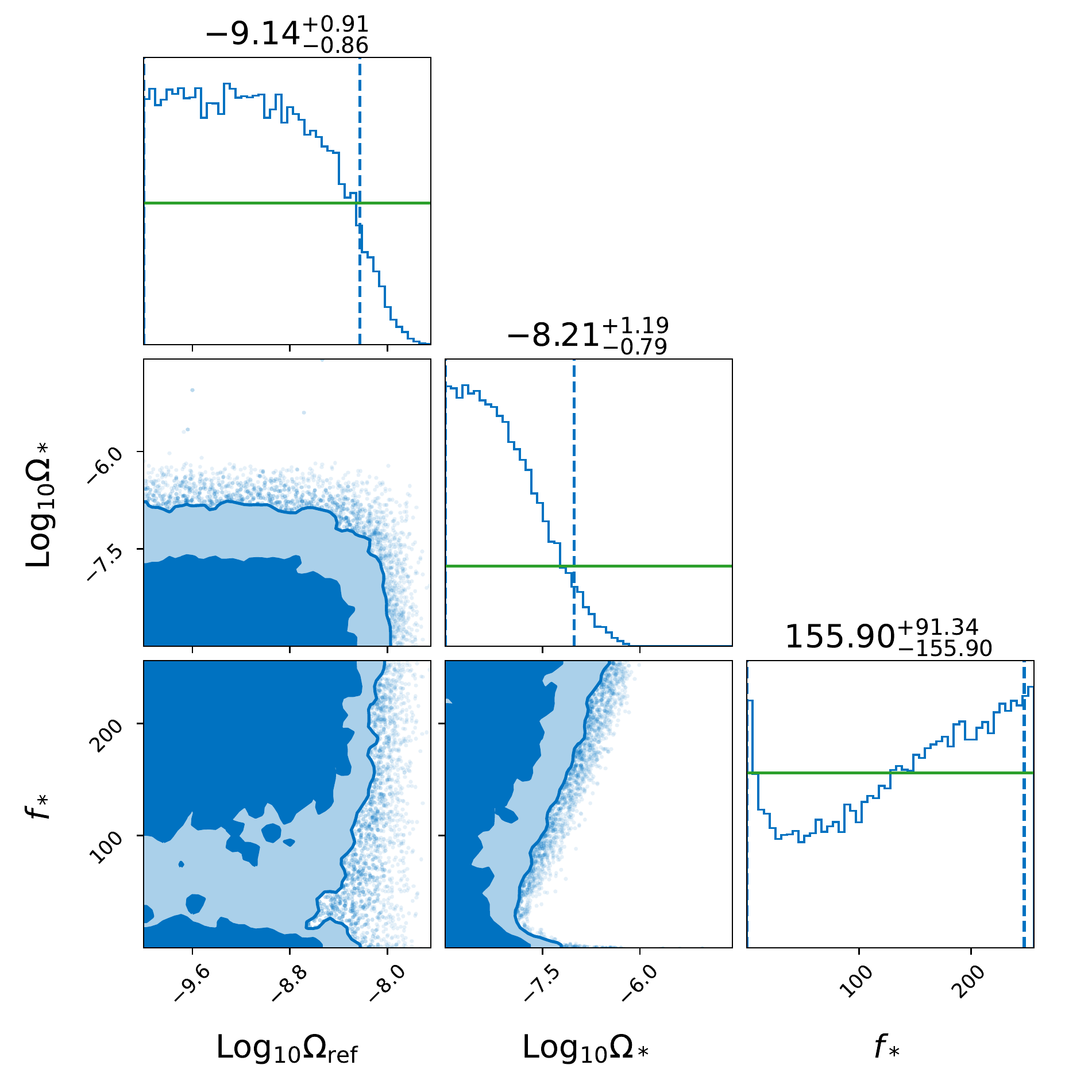}
    \includegraphics[width=0.49\textwidth]{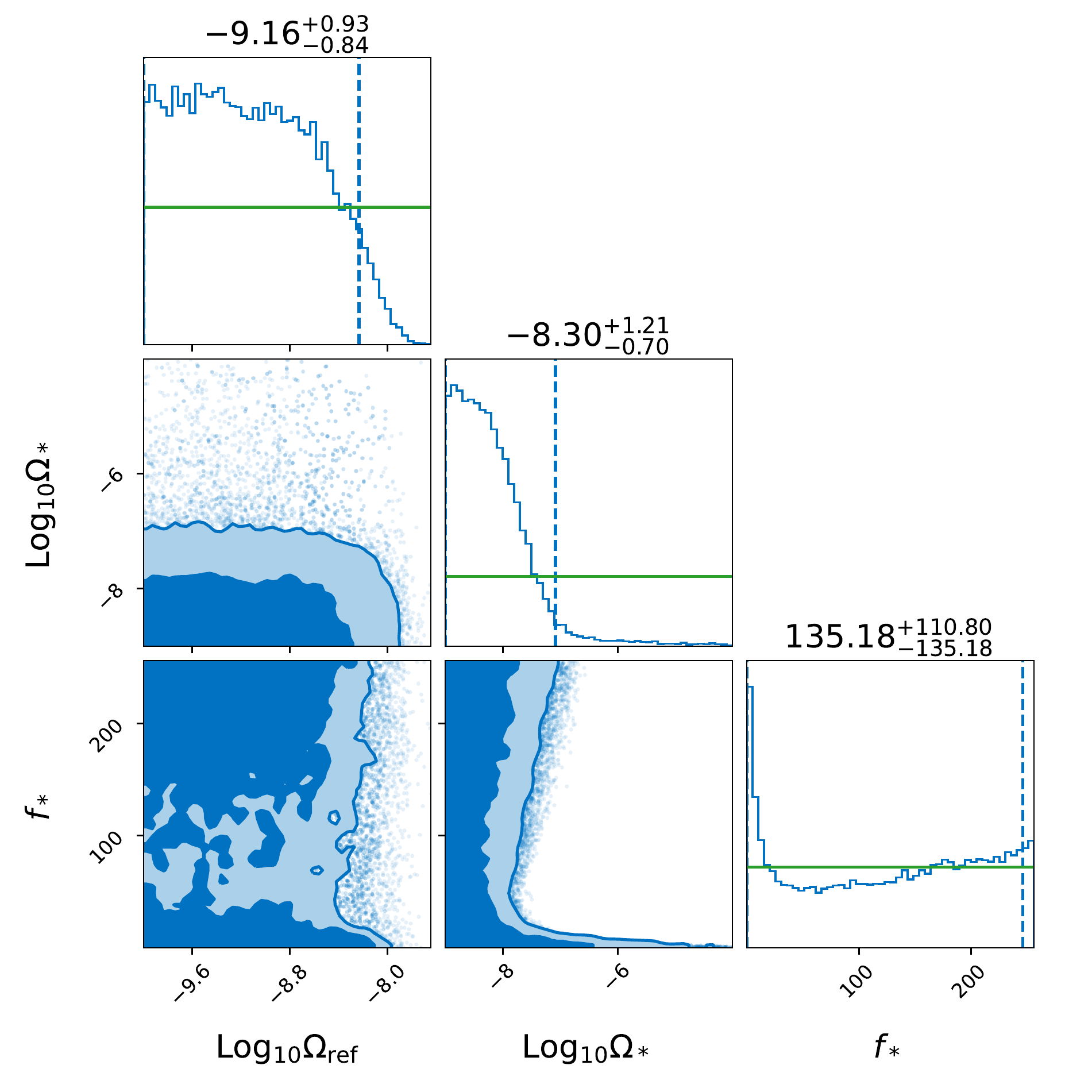}\\
    \caption{The parameter posteriors for the BPL+CBC model with BPL priors fixed by envelope collisions (left) and  sound waves (right). The $68\%$ and $95\%$ contours are depicted in colors. The vertical dashed lines describe the $95\%$ and the horizon lines are the prior distributions. We also show the (0, 95\%) of parameters individually.}
    \label{fig:BPL.pdf}
\end{figure*}

In fact, with fixed $n_1=3, n_2 = -1$ ($n_2 = -4$), and $\Delta=4$ ($\Delta = 2$) for SGWBs from envelope collisions (sound waves), the $95\%$ upper limit we found on the peak amplitude reads $\Omega_*<9.7\times10^{-8}$ ($\Omega_*<8.2\times10^{-8}$), which, along with posterior sample of $f_*$ from the Fig.~\ref{fig:BPL.pdf}, can be combined into a posterior of $\Omega_\mathrm{BPL}$, leading to a $95\%$ C.L. constraint $\Omega_\mathrm{BPL}(25\,\mathrm{Hz})<3.5 \times 10^{-9}$ ($\Omega_\mathrm{BPL}(25\,\mathrm{Hz})<6.4 \times 10^{-9}$) at the CBC reference frequency $f_*=25$ Hz, while the CBC reference amplitude is found to be bounded by $\Omega_\mathrm{ref}<5.9\times10^{-9}$ ($\Omega_\mathrm{ref}<5.9\times10^{-9}$). The Bayes ratios $\log \mathcal{B}_\mathrm{Noise}^{\mathrm{BPL + CBC}} = - 1.72$ ($\log \mathcal{B}_\mathrm{Noise}^{\mathrm{BPL + CBC}} = - 1.86$) and $\log \mathcal{B}_\mathrm{CBC}^{\mathrm{BPL + CBC}} = -1.19$ ($\log \mathcal{B}_\mathrm{CBC}^{\mathrm{BPL + CBC}} = -1.33$) even more disfavor for a BPL GW spectrum from envelope collisions (sound waves) over SGWBs from either pure Gaussian noises or CBCs than Ref.~\cite{Romero:2021kby}, slightly improving the previous result.


\subsection{The DP+CBC model}



\begin{figure*}
    \centering
    \includegraphics[width=0.6\textwidth]{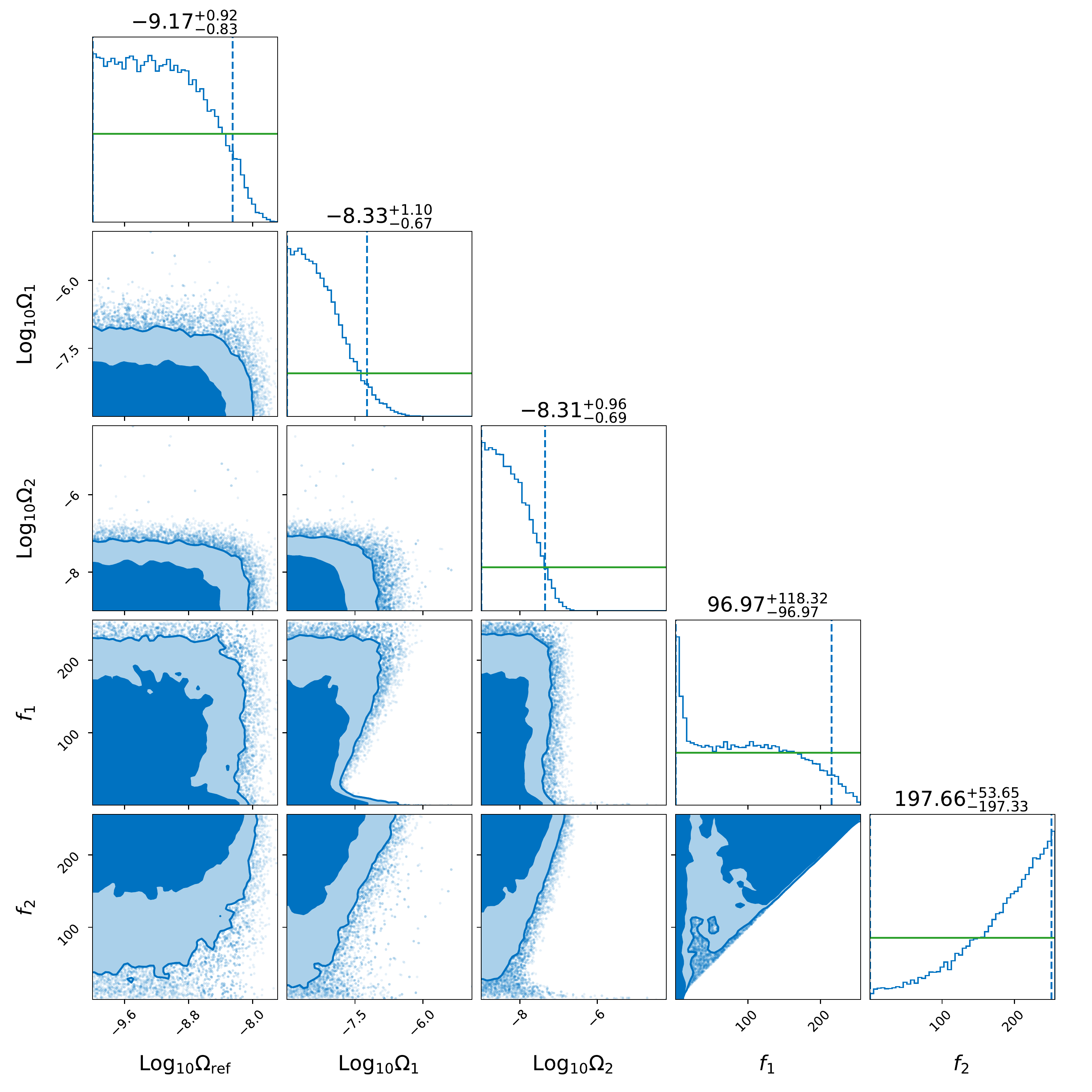}\\
    \caption{The parameter posteriors for the BP (envelope collisions + sound waves) +CBC model. The $68\%$ and $95\%$ contours are depicted in colors. The vertical dashed lines describe the $95\%$ and the horizon lines are the prior distributions. We also show the (0, 95\%) of parameters individually.}
    \label{fig:TBPL}
\end{figure*}

For the SGWBs from FOPTs, the present peak frequency of envelope collisions~\cite{Kosowsky:1992vn,Kosowsky:1992rz,Huber:2008hg,Weir:2016tov,Jinno:2016vai,Cutting:2018tjt,Cutting:2020nla,Gould:2021dpm},
\begin{align}
f_\mathrm{env}&=16.5\left(\frac{f_\mathrm{bc}}{\beta}\right)\left(\frac{\beta}{H_\mathrm{pt}}\right)\left(\frac{T_\mathrm{pt}}{100\,\mathrm{GeV}}\right)\left(\frac{g_*}{100}\right)^\frac16\,\mu\mathrm{Hz}\nonumber\\
&<5.775\left(\frac{\beta}{H_\mathrm{pt}}\right)\left(\frac{T_\mathrm{pt}}{100\,\mathrm{GeV}}\right)\left(\frac{g_*}{100}\right)^\frac16\,\mu\mathrm{Hz},
\end{align}
is always smaller than the present peak frequency of sound waves~\cite{Hindmarsh:2013xza,Hindmarsh:2015qta,Hindmarsh:2017gnf,Cutting:2019zws},
\begin{align}
f_\mathrm{sw}&=\frac{19}{v_w}\left(\frac{\beta}{H_\mathrm{pt}}\right)\left(\frac{T_\mathrm{pt}}{100\,\mathrm{GeV}}\right)\left(\frac{g_*}{100}\right)^\frac16\,\mu\mathrm{Hz}\nonumber\\
&>19\left(\frac{\beta}{H_\mathrm{pt}}\right)\left(\frac{T_\mathrm{pt}}{100\,\mathrm{GeV}}\right)\left(\frac{g_*}{100}\right)^\frac16\,\mu\mathrm{Hz},
\end{align}
for the bubble-wall velocity $0<v_w<1$, where $f_\mathrm{bc}=0.35\beta/(1+0.069v_w+0.69v_w^4)<0.35\beta$ is the peak frequency of bubble collisions right after the phase transition, $\beta/H_\mathrm{pt}$ is the Hubble time scale $H_\mathrm{pt}^{-1}$ relative to the PT duration $\beta^{-1}$ at the PT temperature $T_\mathrm{pt}$, and $g_*$ is the effective number of relativistic degrees of freedom. Therefore, we can specifically fix the priors $f_{*,1}<f_{*,2}$ with $n_{2,1} = -1, \Delta_1=4$ (envelope collisions) and $n_{2,2} = -4, \Delta_2=2$ (sound waves) as well as $n_{1,i}=3$ (by causality) for both $i=1$ (envelope collisions) and $i=2$ (sound waves), and then place $95\%$ C.L. upper limits on the low-frequency peak amplitude $\Omega_{*,1}<5.8\times10^{-8}$ and high-frequency peak amplitude $\Omega_{*,2}<4.4\times10^{-8}$, while the CBC reference amplitude is bounded by $\Omega_\mathrm{ref}<5.6\times10^{-9}$ as obtained from Fig.~\ref{fig:TBPL}. The Bayes ratios $\log\mathcal{B}_\mathrm{Noise}^\mathrm{DP+CBC}=-3.25$ and $\log\mathcal{B}_\mathrm{CBC}^\mathrm{DP+CBC}=-2.72$ strongly disfavor for the DP+CBC over either noises or CBCs than the single BPL+CBC model does.

\begin{figure}
    \centering
    \includegraphics[width=0.45\textwidth]{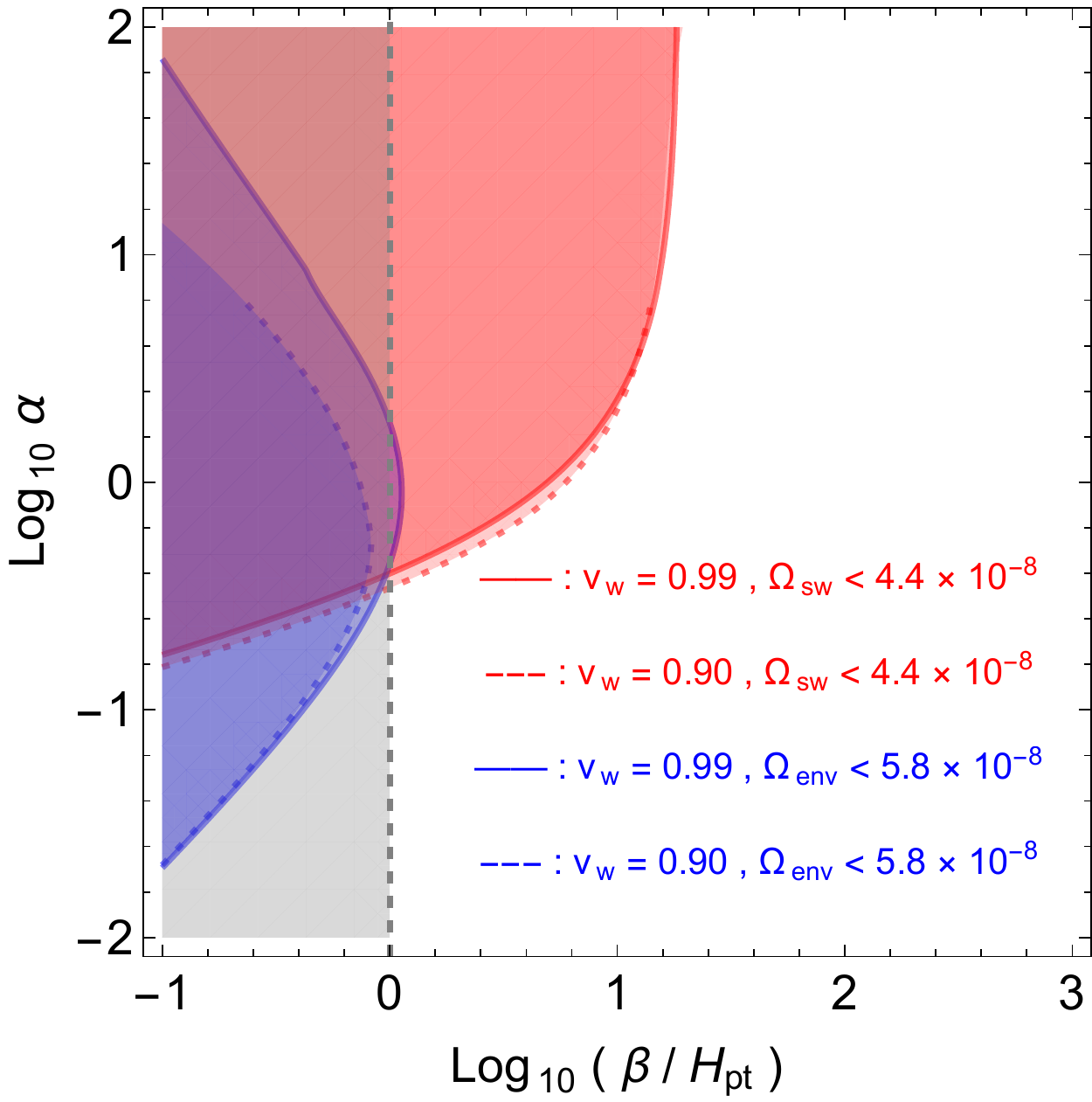}
    \caption{The implied constraints on the FOPT parameters $\beta/H_\mathrm{pt}$ and $\alpha$ for the bubble-wall velocities $v_w=0.9$ (dashed) and $v_w=0.99$ (solid) from the constraints on the DP+CBC model. The blue and red shaded regions are ruled out by the upper bounds on the low-frequency and high-frequency peak amplitudes, respectively. The gray-shaded region is usually not considered for the FOPT to complete successfully.}
    \label{fig:FOPTTBPL}
\end{figure}

The above constraints on the low-frequency and high-frequency peak amplitudes can be transformed into constraints on the PT inverse duration $\beta/H_\mathrm{pt}$ and  strength factor $\alpha$ for a given bubble-wall velocity $v_w$. The peak amplitude of envelope collisions is known as~\cite{Kosowsky:1992vn,Kosowsky:1992rz,Huber:2008hg,Weir:2016tov,Jinno:2016vai,Cutting:2018tjt,Cutting:2020nla,Gould:2021dpm}
\begin{align}
\Omega_\mathrm{env}=1.67\times10^{-5}\frac{A}{h^2}\left(\frac{H_\mathrm{pt}}{\beta}\right)^2\left(\frac{\kappa_\phi\alpha}{1+\alpha}\right)^2\left(\frac{100}{g_*}\right)^\frac13,
\end{align}
where $A(v_w)\equiv0.48v_w^3/(1+5.3v_w^2+5v_w^4)$ is the amplitude and $\kappa_\phi$ is the efficiency factor of inserting released vacuum energy into the bubble wall kinetic energy evaluated generally in Ref.~\cite{Cai:2020djd}. For the most optimistic constraint, we can take a crude estimation $\kappa_\phi\approx1-\kappa_\mathrm{sw}$ from the efficiency factor $\kappa_\mathrm{sw}$ of fluid motions given shortly below. The peak amplitude of sound waves is known as~\cite{Hindmarsh:2013xza,Hindmarsh:2015qta,Hindmarsh:2017gnf,Cutting:2019zws}
\begin{align}
\Omega_\mathrm{sw}=2.65\times10^{-6}\frac{v_w}{h^2}\left(\frac{H_\mathrm{pt}}{\beta}\right)\left(\frac{\kappa_\mathrm{sw}\alpha}{1+\alpha}\right)^2\left(\frac{100}{g_*}\right)^\frac13\Upsilon,
\end{align}
where the efficiency factor $\kappa_\mathrm{sw}(\alpha,v_w)$ of bulk fluid motions can be fitted as a function of $\alpha$ and $v_w$ by hydrodynamics~\cite{Espinosa:2010hh} (see also~\cite{Wang:2022lyd} for the varying sound velocity generalization to the constant sound velocity estimations~\cite{Giese:2020rtr,Giese:2020znk,Wang:2020nzm}). The suppression factor $\Upsilon\equiv1-(1+2\tau_\mathrm{sw}H_\mathrm{pt})^{-1/2}$~\cite{Guo:2020grp} accounts for the finite lifetime of sound waves from the onset timescale of turbulence, $\tau_\mathrm{sw}H_\mathrm{pt}\approx(8\pi)^{1/3}v_w/(\beta/H_\mathrm{pt})/\bar{U}_f$, with the root-mean-squared fluid velocity given by $\bar{U}_f^2=3\kappa_\mathrm{sw}\alpha/[4(1+\alpha)]$~\cite{Weir:2017wfa}. In all cases, we can take the effective number of degrees of freedom $g_*=100$ and dimensionless Hubble constant $h=0.67$ for illustration. Therefore, both peak amplitudes $\Omega_\mathrm{env}$ and $\Omega_\mathrm{sw}$ can be expressed for a given $v_w$ as functions of $\beta/H_\mathrm{pt}$ and $\alpha$, which can be further constrained in Fig.~\ref{fig:FOPTTBPL} with the blue and red shaded regions ruled out by $\Omega_\mathrm{env}=\Omega_{*,1}<5.8\times10^{-8}$ and $\Omega_\mathrm{sw}=\Omega_{*,2}<4.4\times10^{-8}$, respectively. Although the current GW data cannot put strong constraints on the FOPTs, the very strong FOPTs of super-cooling type in the LIGO-Virgo band with $\alpha\gtrsim\mathcal{O}(1)$ and $\beta/H_\mathrm{pt}\lesssim\mathcal{O}(10)$ can be marginally ruled out from Fig.~\ref{fig:FOPTTBPL}. Note here that there is no precise but conventional definition~\cite{Wang:2020jrd} for the very strong FOPT of super-cooling type. The strength factor $\alpha$ measures the relative size of released vacuum energy density with respect to the background radiation energy density, and hence $\alpha\gtrsim\mathcal{O}(1)$ indicates a very strong FOPT. The other parameter $\beta/H_\mathrm{pt}$ measures the relative size of Hubble horizon scale $H_\mathrm{pt}^{-1}$ with respect to the mean bubble separation $(8\pi)^{1/3}v_w\beta^{-1}\sim\beta^{-1}$, and hence  $\beta/H_\mathrm{pt}\lesssim\mathcal{O}(10)$ indicates a relatively large radius of bubbles at collisions, which would  result in a relatively long PT duration~\cite{Cai:2017tmh,Ellis:2018mja} that leads to ultra-low temperature at percolations than the critical/nucleation temperature (hence the name supercooling).

\subsection{The DB+CBC model} 

\begin{figure*}
    \centering
    \includegraphics[width=0.6\textwidth]{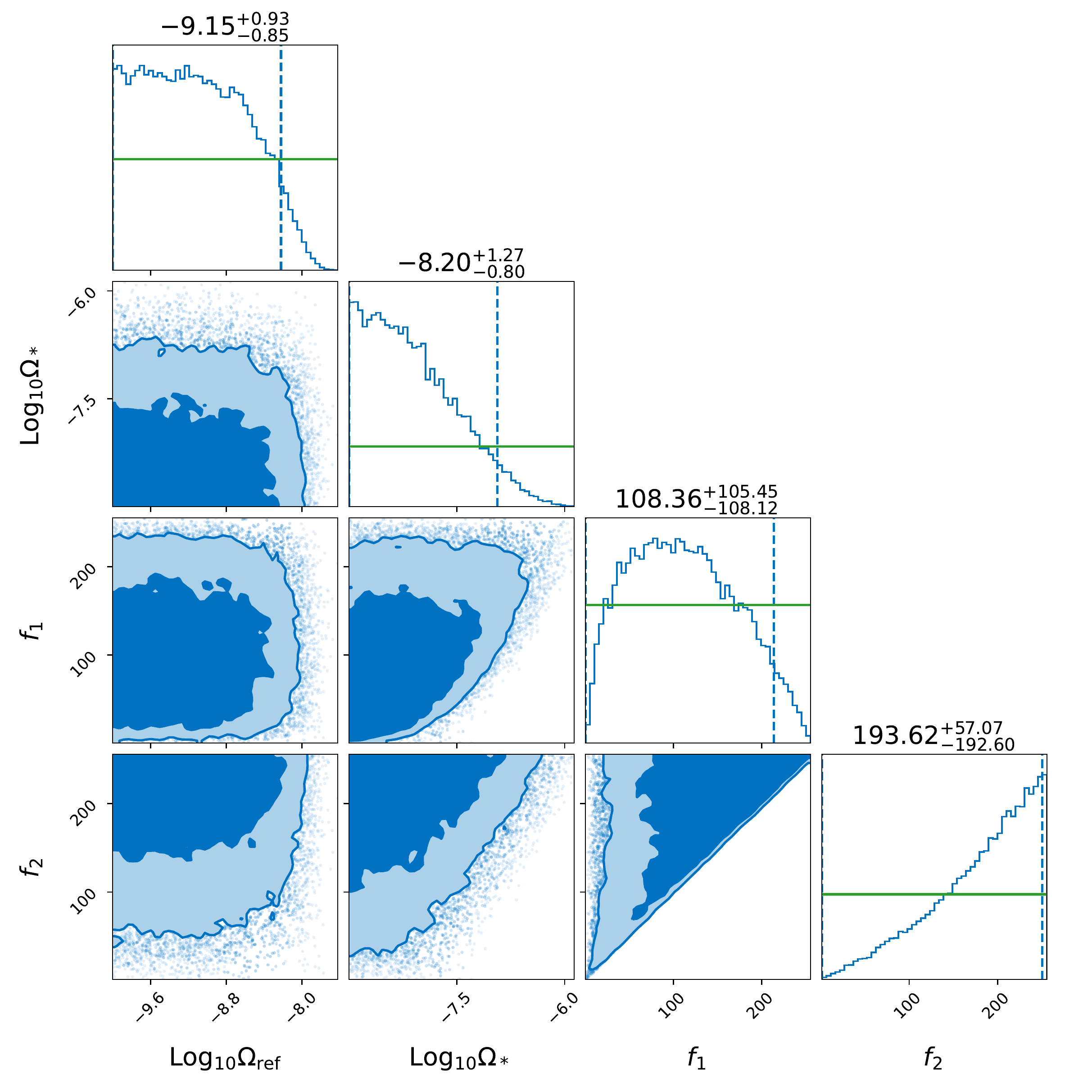}\\
    \caption{The parameter posteriors for the DB (beyond envelope)+CBC model. The $68\%$ and $95\%$ contours are depicted in colors. The vertical dashed lines describe the $95\%$ and the horizon lines are the prior distributions. We also show the (0, 95\%) of parameters individually.}
    \label{fig:DBPL}
\end{figure*}

For a physical process associated with two characteristic length scales, the generated SGWBs usually admit a doubly broken (DB) power-law spectrum. One such example is the cosmological FOPT with the vacuum-bubble collisions characterized by the averaged initial bubble separation and bubble-wall thickness, and sound waves characterized by the averaged initial bubble separation and sound shell thickness~\cite{Hindmarsh:2016lnk,Hindmarsh:2019phv,Jinno:2020eqg,Cai:2023guc}. We consider specifically in this section the GWs from the bubble-wall collisions beyond the envelope approximation with $n_l = 3$, $n_m = 1$, and $n_h = -1$. The overall amplitude can be constrained as $\Omega_*<1.2\times10^{-7}$, which, after combined with the posterior samples of $f_l$, and $f_h$ in Fig.~\ref{fig:DBPL}, renders $95\%$ C.L. upper bound $\Omega_\mathrm{DB}(25\,\mathrm{Hz})<2.3\times10^{-9}$ at the CBC reference frequency with the corresponding CBC reference amplitude bounded by $\Omega_\mathrm{ref}<6.0\times10^{-9}$. Similar to the single BPL+CBC model, the Bayes ratios $\log\mathcal{B}_\mathrm{Noise}^\mathrm{DBPL+CBC}=-1.86$ and $\log\mathcal{B}_\mathrm{CBC}^\mathrm{DB+CBC}=-1.33$ also slightly disfavor for the DB+CBC over either noises or CBCs.

\section{Conclusions and discussions}\label{sec:ConDis}

In this paper, we have implemented the Bayes search for the SGWBs specifically from the cosmological first-order phase transitions with a DP or DB spectrum in the first three observing runs of the Advanced LIGO-Virgo collaborations. No positive evidence has been found for both DP+CBC and DB+CBC models with respective to the backgrounds from either Gaussian noises or CBCs, though the DP+CBC is even more disfavored than the DB+CBC model as well as the usual BPL+CBC model. In particular, our results for the BPL+CBC model slightly improve the previous claim on the null detection for the BPL spectrum, and  the DP+CBC results motivated from FOPTs could marginally rule out the very strong FOPT of super-cooling type in the LIGO-Virgo band. All these results could be further improved for the upcoming fourth observing run of the LIGO/Virgo/KAGRA Collaboration, but currently we are still on the way to uncover the SGWBs.\\

\begin{acknowledgments}
We thank Huai-Ke Guo, Katarina Martinovic, and Alba Romero for fruitful correspondence on the calibration uncertainties in the LIGO/VIRGO analysis. We thank the helpful discussion on the code with Wen-Hong Ruan, Chang Liu and He Wang.
This work is supported by the National Key Research and Development Program of China Grants  No. 2021YFC2203004, No. 2021YFA0718304, and No. 2020YFC2201501,
the National Natural Science Foundation of China Grants No. 12105344, No.12235019, and No. 12047503, 
the Key Research Program of the Chinese Academy of Sciences (CAS) Grant No. XDPB15, the Key Research Program of Frontier Sciences of CAS, 
and the Science Research Grants from the China Manned Space Project No. CMS-CSST-2021-B01.
We also acknowledge the use of the HPC Cluster of ITP-CAS.
\end{acknowledgments}


\bibliography{ref}

\end{document}